\documentclass[apj]{emulateapj}

\usepackage{apjfonts}

\newcommand{\ofour}{[\textrm{O}~\textsc{iv}]~}
\newcommand{\oiv}{[\textrm{O}~\textsc{iv}]}
\newcommand{\othree}{[\textrm{O}~\textsc{iii}]~}
\newcommand{\oiii}{[\textrm{O}~\textsc{iii}]}

\newcommand{\neii}{[\textrm{Ne}~\textsc{ii}]}
\newcommand{\fetwo}{[\textrm{Fe}~\textsc{ii}]~}
\newcommand{\feii}{[\textrm{Fe}~\textsc{ii}]}
\newcommand{\lrest}{\lambda_{\scriptsize{\textnormal{rest}}}}

\shorttitle{AGN Luminosity Indicators}
\shortauthors{Diamond-Stanic et al.}

\begin{document}
\slugcomment{Published in The Astrophysical Journal}

\title{Isotropic Luminosity Indicators in a Complete AGN Sample}

\author{Aleksandar M. Diamond-Stanic\altaffilmark{1}, George
H. Rieke\altaffilmark{1}, Jane R. Rigby\altaffilmark{2}}

\altaffiltext{1}{Steward Observatory, University of Arizona, 933 North
Cherry Avenue, Tucson, AZ, 85721, USA; adiamond@as.arizona.edu}
\altaffiltext{2}{Observatories, Carnegie Institution of Washington,
813 Santa Barbara Street, Pasadena, CA 91101, USA}

\begin{abstract}
The \oiv~$\lambda25.89~\mu$m line has been shown to be an accurate
indicator of active galactic nucleus (AGN) intrinsic luminosity in
that it correlates well with hard (10--200~keV) X-ray emission.  We
present measurements of \ofour for 89 Seyfert galaxies from the
unbiased revised Shapley--Ames (RSA) sample.  The \ofour luminosity
distributions of obscured and unobscured Seyferts are
indistinguishable, indicating that their intrinsic AGN luminosities
are quite similar and that the RSA sample is well suited for tests of
the unified model.  In addition, we analyze several commonly used
proxies for AGN luminosity, including \oiii~$\lambda5007$~\AA, 6~cm
radio, and 2--10~keV X-ray emission.  We find that the radio
luminosity distributions of obscured and unobscured AGNs show no
significant difference, indicating that radio luminosity is a useful
isotropic luminosity indicator.  However, the observed \othree and
2--10~keV luminosities are systematically smaller for obscured
Seyferts, indicating that they are not emitted isotropically.
\end{abstract}


\section{Introduction}

Many differences among active galactic nuclei (AGNs) are explained in
terms of the line of sight to the supermassive black hole, such that
an object will be classified as unobscured (type~1) if the central
continuum source and broad-line region are directly visible, or as
obscured (type~2) if large amounts of gas and dust block the central
region.  Unification schemes \citep[e.g.,][]{ant93} often invoke
obscuring material in a torus geometry, such that the observed
spectral energy distribution depends solely on viewing angle and the
covering fraction of the torus sets the ratio of obscured to
unobscured objects.  This paradigm has been challenged by suggestions
that the obscured-to-unobscured ratio varies as a function of
luminosity \citep[e.g.,][]{ued03,ste03,laf05} and that some
low-luminosity AGNs may not have broad-line regions at all
\citep[e.g.,][]{tra03,bia08,bri08}.

To test models for the geometry of the obscuring material and the
fundamental differences between type 1 and type 2 AGNs, one needs an
unbiased, well-understood sample of objects that includes both
low-luminosity and highly obscured sources.  The spectroscopically
selected, galaxy-magnitude-limited sample drawn from the revised
Shapley--Ames catalog \citep[RSA;][]{sha32,san87} meets these criteria
\citep{mai95, ho97}, and is well suited to probe basic predictions of
AGN behavior.  For example, if the sample is truly unbiased and the
unified model is correct, the intrinsic AGN properties of the obscured
and unobscured members should be the same.

In this paper, we consider 89 Seyferts from \citet{mai95} and
\citet{ho97} drawn from the parent sample of galaxies with
$B_T\leq13$.\footnote{We do not include two galaxies classified as
Seyferts by \citet{ho97}, NGC185 and NGC676.  The former is a dwarf
spheroidal galaxy without a well-defined nucleus \citep{ho95,ho01},
and the latter is contaminated by a bright star 5$\arcsec$ from its
nucleus.}  This sample (see Table~\ref{table:data}) includes 18
Seyfert 1s (type 1.0--1.5, hereafter Sy1s) and 71 Seyfert 2s (type
1.8--2, hereafter Sy2s).  We use it to probe whether there is a
systematic luminosity difference between obscured and unobscured AGNs.
Such a difference would be expected if AGN obscuration were luminosity
dependent \citep[e.g.,][]{law91} or if there existed a significant
population of low-luminosity AGNs that lack a broad-line region
\citep[e.g.,][]{lao03,nic03} in the sense that Sy2s would be
disproportionately represented at faint luminosities.  We determine
the AGN luminosity through measurements of the \ofour emission line at
$25.89~\mu$m (ionization potential 54.9~eV, critical density
$10^4$~cm$^{-3}$), which has been established as an accurate
luminosity indicator by \citet{mel08a} and \citet{rig09} by comparison
to hard ($E>10$~keV) X-rays.  We also compile measurements from the
literature of quantities that are thought to be luminosity indicators,
including \oiii~$\lambda5007$~\AA, 2--10~keV X-ray, and 6~cm radio
emission, to determine which are in fact isotropically emitted.

\section{Data}\label{sec:data}

We gather data from the {\it Spitzer Space Telescope} \citep{wer04}
archive taken with the Infrared Spectrograph \citep[IRS;][]{hou04} in
the first order of the Long-Low module (LL1;
$\lambda=19.5$--$38.0~\mu$m).  The slit size for this order is
$10.7\arcsec\times168\arcsec$ and the resolution is $R=64$--$128$.
For data taken in staring mode, we begin our analysis on the
post--basic calibrated data produced by the {\it Spitzer} Science
Center pipeline and compute a weighted average of the one-dimensional
spectra extracted at each of the nod positions.  For data taken in
mapping mode, we begin our analysis with the basic calibrated data and
use the CUBISM software \citep{smi07} to combine two-dimensional
images and extract one-dimensional spectra.  To obtain flux
calibration appropriate for point sources, we disable the FLUXCON and
SLCF options within CUBISM, and use $10.7\times35.2\arcsec$ apertures
centered on the nucleus of the galaxy; this aperture corresponds to
the LL1 slit size ($10.7\arcsec$) and the default point-source
extraction aperture size at $26~\mu$m ($35.2\arcsec$).

For each spectrum, we fit a power law to the continuum using the
rest-frame wavelength regions $24.75$--$25.5~\mu$m and
$26.5$--$27~\mu$m.  We then fit a Gaussian to the \ofour line and
calculate the error on the flux measurement using the uncertainty in
the five pixels closest to $\lrest=25.89~\mu$m and the rms of the
continuum fit in the wavelength regions mentioned above.  For cases
where this method yields a $<5\sigma$ line detection, we inspect the
spectrum visually to determine whether the line is confidently
detected.  If there is not a clear detection, we calculate a
conservative upper limit by adding $3\sigma$ to the best-fit flux.
The IRS LL1 data for NGC1068, CIRCINUS, and NGC4945 are saturated, so
we take fluxes from ISO-SWS spectra published by \citet{stu02} and
\citet{spo00}.

A source of uncertainty for \ofour fluxes measured from low-resolution
IRS spectra is contamination by \feii~$\lambda25.99~\mu$m (ionization
potential 7.9~eV) emission associated with star formation.  Spectra
from the IRS Long-High module (LH, $R\sim600$) are available for 68/70
of the Seyferts with an LL1 line detection, so we are able to measure
the amount of \fetwo contamination.  We analyze the post--basic
calibrated data from LH order 15 ($\lambda=25.0-27.4~\mu$m) and fit a
Gaussian to each of the two lines.  For sources with LL1 equivalent
widths (EWs) greater than 0.10~$\mu$m, the median value for the
sample, we find that the \fetwo contribution is small ($<15\%$ in all
cases).  Among the sources with lower EWs, most still have $<25\%$
\fetwo contributions, but a few are actually dominated by \fetwo
(NGC3079, NGC4579, NGC4594, NGC5005).  We apply a correction to the
LL1 \ofour measurements for all sources that have an LH \fetwo
detection.  The \ofour fluxes and uncertainties are listed in
Table~\ref{table:data}.

For the purpose of determining whether sources with large \fetwo
contributions can be identified without high-resolution data, we
consider the \neii~$\lambda12.81~\mu$m (ionization potential 21.6~eV)
emission, which is also associated with star formation.  Inspection of
IRS data from the first order of the Short-Low module
($\lambda=7.4-14.5~\mu$m) indicates that sources with $>15\%$ \fetwo
contributions also have strong \neii\ lines; the ratio of \neii\ to
\oiv+\fetwo for these sources is always unity or greater.  Indeed, all
19 sources with $>15\%$ \fetwo contributions can be identified as
having both $\ofour\textnormal{EW}\leq0.1~\mu$m and $\neii/\oiv\geq1$.
There are, however, an additional eight sources that meet these
criteria, but only have $\sim10\%$ \fetwo contributions.

Utilizing the NED\footnote{http://nedwww.ipac.caltech.edu/.} and
HEASARC\footnote{http://heasarc.gsfc.nasa.gov/.} databases, we searched
the literature to gather \oiii~$\lambda5007$~\AA\ emission-line
fluxes, 2--10~keV X-ray fluxes, and 6~cm radio flux densities.  These
values and the corresponding references are included in
Table~\ref{table:data}.  When multiple values were available, we gave
preference to measurements with smaller beam sizes that isolate the
nuclear emission from that of the host galaxy.  All values are
observed quantities that have not been corrected for extinction.  For
the galaxies with extinction-corrected \othree fluxes published by
\citet{vac97} and \citet{win92}, we calculated observed \othree fluxes
based on the assumed dust reddening.  All sources have published 6~cm
flux densities or upper limits, and all except NGC4945 have published
\othree fluxes.  The X-ray coverage of the sample is less complete,
but 72/89 galaxies have published 2--10~keV fluxes, and an additional
nine sources have unpublished XMM-Newton archival data.  For these
nine sources, we use European Photon Imaging Camera count rates and
flux measurements in the 2.0--4.5~keV and 4.5--12.0~keV bands from the
XMM-Newton Serendipitous Source Catalogue \citep{wat09} along with the
power-law photon index $\Gamma$ inferred using PIMMS
v3.9i\footnote{http://heasarc.nasa.gov/Tools/w3pimms.html.} to estimate
2--10~keV fluxes.  More complete X-ray spectral analysis for these
sources is deferred to future work.\footnote{However, inspection of
pipeline products from XMM-Newton Science Archive Version 5.0
(http://xmm.esac.esa.int/xsa/) reveals a strong
($\textnormal{EW}>1$~keV) Fe K$\alpha$ emission line in the spectrum
of NGC7479, indicative of a Compton-thick source.}

For galaxies studied by \citet{ho97}, we use distances from their
Table 10 with exceptions for NGC1058, NGC3031, NGC4258, NGC4395, and
NGC5194 (see Table~\ref{table:data}).  For the remaining galaxies we
use distances from NED that are calculated assuming
$H_0=73$~km$^{-1}$~s$^{-1}$~Mpc$^{-1}$ and velocity-field corrected
using the \citet{mou00} model, which includes the influence of the
Virgo cluster, the Great Attractor, and the Shapley supercluster.

\begin{figure}
\begin{center}
\includegraphics[angle=0,scale=.45]{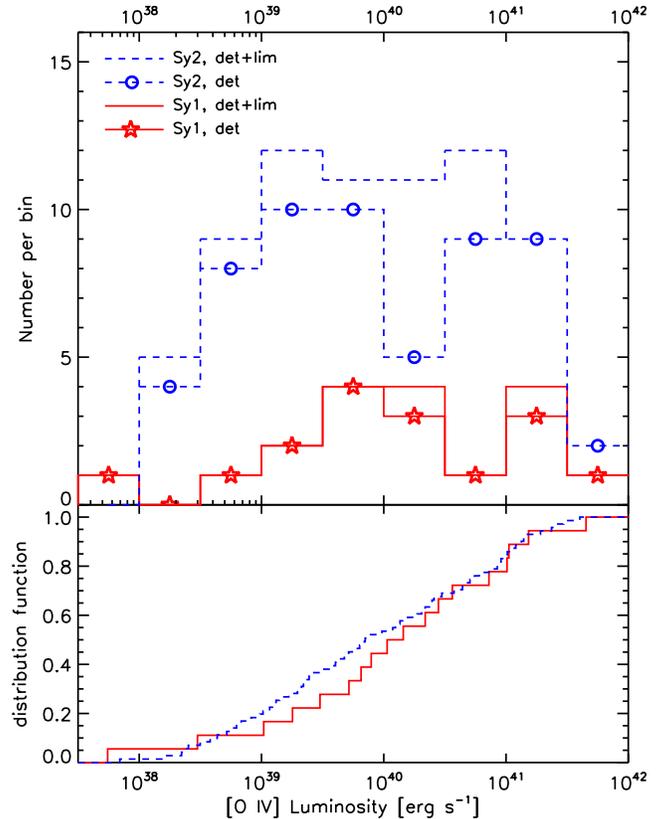}
\caption{The distribution of \oiv~$\lambda25.89~\mu$m luminosities for
Seyfert galaxies in the RSA sample.  The top panel shows histograms
for Sy2 detections and upper limits (dashed blue line), Sy2 detections
(dashed blue line marked by blue circles), Sy1 detections and upper
limits (solid red line), and Sy1 detections (solid red line marked by
red stars).  The bottom panel shows the empirical distribution
functions for Sy2s and Sy1s.  The distributions are not statistically
distinguishable.}
\label{fig:o4}
\end{center}
\end{figure}

\begin{figure}
\begin{center}
\includegraphics[angle=0,scale=.4]{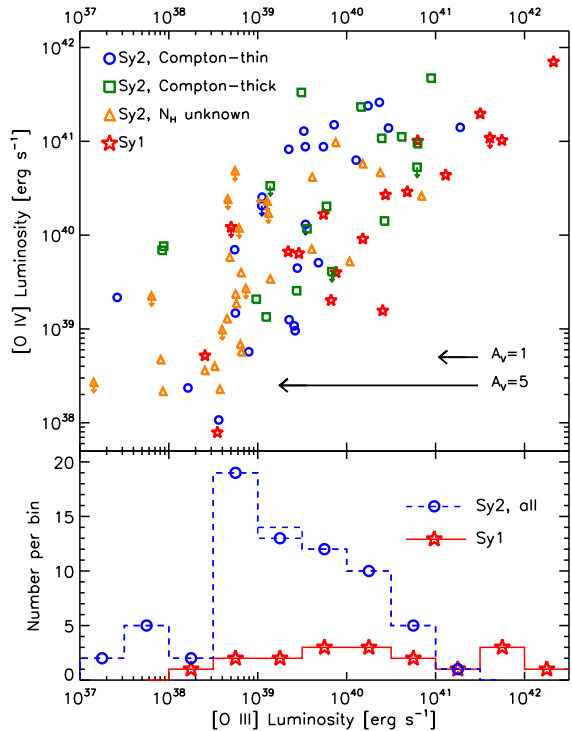}
\caption{Top panel: The relationship between \oiv~$\lambda25.89~\mu$m
and \oiii~$\lambda5007$~\AA\ luminosities.  No extinction corrections
have been applied.  Sy1s are indicated by red stars, Compton-thin
($N_H<10^{24}$~cm$^{-2}$) Sy2s are indicated blue circles,
Compton-thick ($N_H>10^{24}$~cm$^{-2}$) Sy2s are indicated by green
squares, and Sy2s with unknown column densities are indicated by
orange triangles.  While most Sy1s have order unity \oiv/\othree
luminosity ratios, a number of Sy2s are significantly brighter in
\ofour by up to a factor of $\sim100$.  We interpret this as being due
to large host galaxy obscuration in some Sy2s.  The arrows indicate
the effects of extinction on the observed \othree luminosity for
$A_V=1$ and $A_V=5$.  Bottom panel: the distribution of \othree
luminosities, with histogram colors and symbols as in
Figure~\ref{fig:o4}.  The distributions for Sy1s and Sy2s are
statistically different with $p<0.005$.}
\label{fig:o3}
\end{center}
\end{figure}

\begin{figure}
\begin{center}
\includegraphics[angle=0,scale=.4]{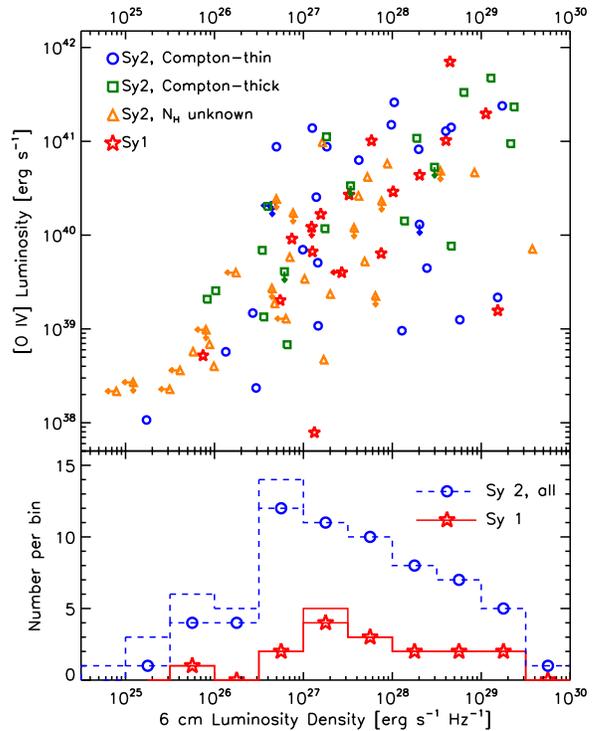}
\caption{Top panel: The relationship between \oiv~$\lambda25.89~\mu$m
and 6~cm radio luminosities.  The symbols are as in
Figure~\ref{fig:o3}.  There is large scatter in this relationship, but
also significant overlap between the various Seyfert types.  Bottom
panel: The distribution of 6~cm luminosities, with histogram colors
and symbols as in Figure~\ref{fig:o4}.  The distributions for Sy1s and
Sy2s are not statistically distinguishable.}
\label{fig:radio}
\end{center}
\end{figure}

\begin{figure}
\begin{center}
\includegraphics[angle=0,scale=.4]{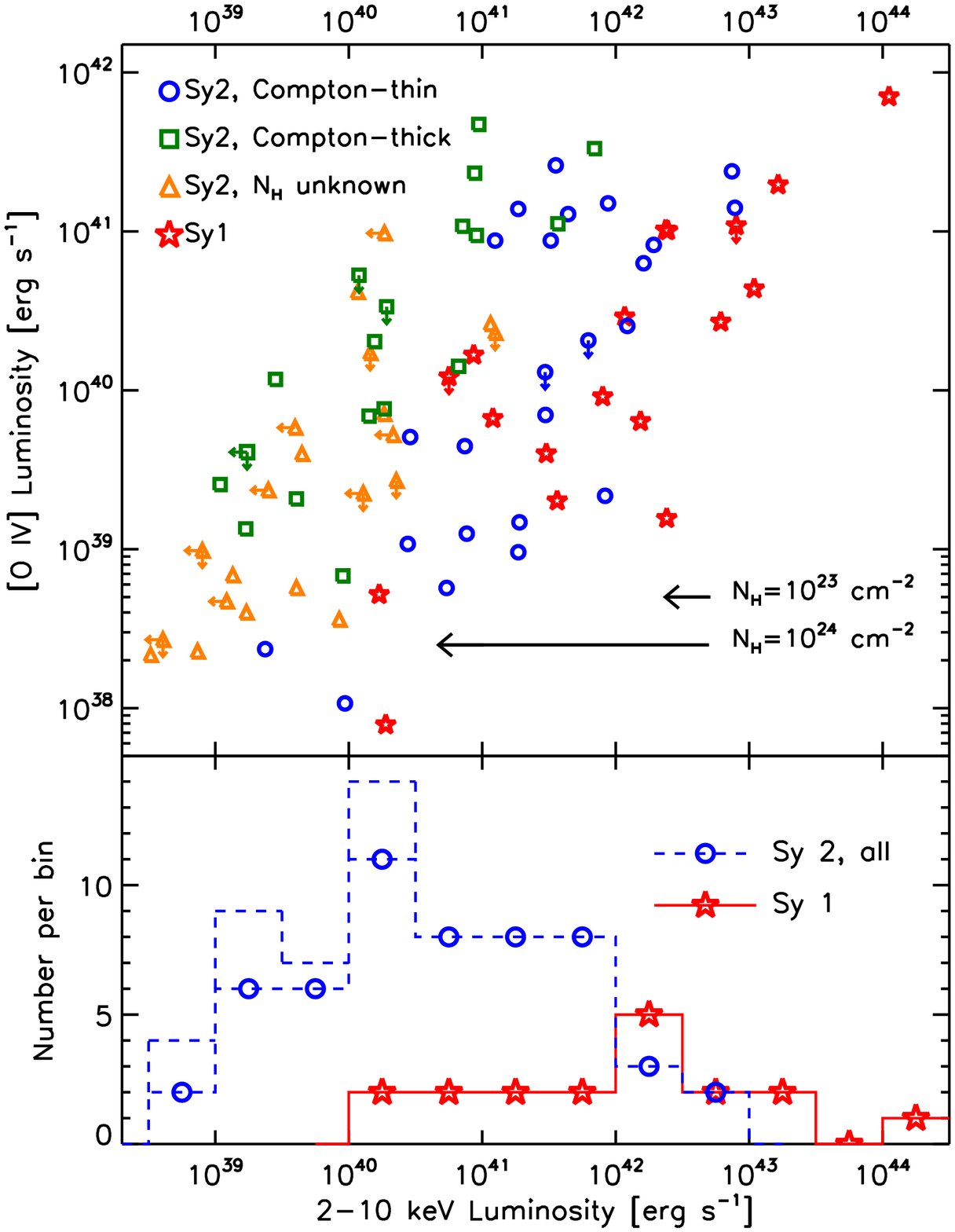}
\caption{Top panel: The relationship between observed
\oiv~$\lambda25.89~\mu$m and 2--10~keV X-ray luminosities.  No
extinction corrections have been applied.  The symbols are as in
Figure~\ref{fig:o3}.  A clear sequence can be seen between the
location of Sy1s, Compton-thin Sy2s, and Compton-thick Sy2s that
reflects increasing amounts of X-ray obscuration.  The observed
2--10~keV X-ray emission is not a reliable indicator of AGN power for
sources with significant obscuration.  The arrows indicate the effect
of column density $N_H=10^{23}$~cm$^{-2}$ and $N_H=10^{24}$~cm$^{-2}$
on the observed X-ray luminosity.  Bottom panel: The distribution of
2--10~keV luminosities, with histogram colors and symbols as in
Figure~\ref{fig:o4}.  The distributions for Sy1s and Sy2s are
statistically different with $p<1\times10^{-5}$.  This figure
illustrates the strong bias against Sy2 galaxies, and particularly
against Compton-thick Sy2s in X-ray-selected AGN samples.}
\label{fig:xray}
\end{center}
\end{figure}

\section{Comparison of Isotropic AGN Indicators}

Commonly proposed isotropic indicators of AGN luminosity include
\oiii~$\lambda5007$~\AA, radio, and hard X-ray emission.  Our
\oiv~$\lambda25.89~\mu$m measurements and the data we have gathered
from the literature let us compare these indicators in obscured and
unobscured members of the RSA Seyfert sample.

We present the distribution of \ofour luminosities in
Figure~\ref{fig:o4}; the distributions for Sy2s and Sy1s are quite
similar.  We utilize two-sample statistical tests that take upper
limits into account \citep{fei85}, and find that the two samples are
consistent with being drawn from the same parent distribution (see
Table~\ref{table:oiv_stats}).  When the Sy2s are grouped by X-ray
column density, we find that both Compton-thin
($N_H<10^{24}$~cm$^{-2}$) and Compton-thick ($N_H>10^{24}$~cm$^{-2}$)
Sy2s are statistically indistinguishable from Sy1s.  The only
statistically significant difference is found when comparing the Sy1s
to the Sy2s without published column densities.  This latter group is
biased towards X-ray-faint sources that do not have enough counts for
a column density measurement and probably tend to have lower intrinsic
luminosities.  In Figure~\ref{fig:o4}, the largest deviation between
the Sy1 and Sy2 distributions occurs in the
$L_{\scriptsize{\oiv}}=10^{38.5}$--$10^{40}$~erg~s$^{-1}$ range, where
these ``$N_H$ unknown'' Sy2s are concentrated.  We emphasize that this
deviation does not produce a statistically significant effect in the
overall Sy2 sample, and is less pronounced than the apparent excess of
low-luminosity ($L_{\scriptsize{\oiv}}<10^{40.5}$~erg~s$^{-1}$) Sy2s
relative to Sy1s seen in a hybrid sample of local Seyfert galaxies by
\citet{mel08b}.

The \othree luminosities are presented in Figure~\ref{fig:o3} and the
results of statistical tests are presented in
Table~\ref{table:o3_stats}.  We find a statistically significant
difference between the \othree luminosity distributions of Sy1s and
Sy2s; the probability that the two samples are drawn from the same
parent distribution is $p<0.005$.  While Sy1s tend to have observed
$\oiv/\oiii$ ratios of order unity, a sizable fraction of the Sy2s
have significantly larger ratios (e.g., all 21 objects with both lines
detected and $\oiv/\oiii>5$ are Sy2s; see the upper-left side of
Figure~\ref{fig:o3}).  We interpret this behavior as being due to
larger host galaxy obscuration towards the narrow-line region in Sy2s.
Similarly, \citet{haa05} invoke optical extinction to explain the
higher $\oiv/\oiii$ ratios in FR2 radio galaxies relative to quasars.
The fluxes in Table~\ref{table:data} are not corrected for extinction,
and in some cases the \othree extinction corrections implied by
optical diagnostics such as the Balmer decrement are substantial
\citep[e.g.,][]{bas99}.  However, applying such a correction does not
always yield a satisfactory result --- two of the galaxies with the
largest $\oiv/\oiii$ ratios, NGC3281 and NGC5128, only exhibit
moderate Balmer reddening H$\alpha$/H$\beta\simeq6$.  This corresponds
to extinction by a factor of $\simeq10$ at 5007~\AA, not sufficient to
explain the extreme values $\oiv/\oiii\simeq100$.  This discrepancy
can be explained if \ofour is detected from heavily extincted regions
that are optically thick ($\tau>>1$) at visible wavelengths, while the
\oiii, H$\alpha$, and H$\beta$ lines are detected exclusively from
less-extincted regions, resulting in a shallower observed Balmer
decrement that underestimates the true extinction.  Interestingly,
there is no statistically significant difference between the \othree
distribution of Sy1s and Compton-thick Sy2s, but this is because the
known Compton-thick sources are biased towards high luminosities;
lower-luminosity sources with $N_H>10^{24}$~cm$^{-2}$ likely exist in
the ``$N_H$ unknown'' category, but have not yet been individually
identified as Compton thick.  We conclude that
\oiii~$\lambda5007~$\AA\ is a significantly less reliable quantitative
indicator of AGN activity than is \oiv~$\lambda25.89~\mu$m.

We show in Figure~\ref{fig:radio} and Table~\ref{table:radio_stats}
that the distributions of radio luminosity density for Sy1s and Sy2s
are indistinguishable.  This result is not surprising given that radio
emission is unaffected by dust.  The Sy2s without column density
measurements are shown, once again, to be intrinsically weaker on
average than the rest of the sample.  The only source in the sample
that exceeds the canonical radio luminosity threshold for radio-loud
AGNs \citep[$L_{\nu}>10^{32}$~erg~s$^{-1}$~Hz$^{-1}$; e.g.,][]{mil90}
is NGC1275, but a handful of objects in Figure~\ref{fig:radio} have
large 6~cm/\ofour flux ratios that are suggestive of a
radio-intermediate classification.  Besides NGC1275, which falls
beyond the range plotted in Figure~\ref{fig:radio}, the 10 objects
with the largest ratios are NGC7213, NGC5128, NGC2639, NGC4594,
NGC3031, NGC4579, NGC3079, NGC2655, NGC4168, and NGC4472.  With these
sources excluded, the scatter in Figure~\ref{fig:radio} reduces from
0.93 dex to 0.57 dex.  While this scatter is large, and radio
selection is biased towards sources that emit a larger fraction of
their bolometric luminosity in the radio, we conclude that radio
luminosity is an isotropic AGN indicator.

The observed 2--10~keV X-ray luminosities of Sy2s are biased compared
to those of Sy1s, and with high statistical significance,
$p<1\times10^{-5}$.  This behavior can be seen in
Figure~\ref{fig:xray} and Table~\ref{table:xray_stats}, and is not
surprising given that one expects typical Sy2 gas column densities of
$10^{22}$--$10^{25}$~cm$^{-2}$ to absorb $10\%$--$100\%$ of the flux
in the 2--10~keV energy range.  Only much higher X-ray energies
promise to be accurate measures of AGN activity levels, although the
most obscured sources will still be affected even at $>20$~keV
energies \citep[e.g.,][]{mel08a,rig09}.  There is a clear offset in
Figure~\ref{fig:xray} between the points corresponding to Sy1s,
Compton-thin Sy2s, and Compton-thick Sy2s as one moves towards smaller
observed X-ray luminosities.  The objects with unknown column
densities fall in between and overlap with the Compton-thin and
Compton-thick Sy2s, suggesting that many are highly absorbed and that
a significant fraction is likely to be Compton-thick.  We note that
8/29 objects in this $N_H$ unknown category have no data in the
2--10~keV range, and thus do not appear in Figure~\ref{fig:xray}, nor
are they included in the statistical tests.

\section{Discussion}

We have found that the Sy1s and Sy2s in the RSA sample have quite
similar \oiv~$\lambda25.89~\mu$m luminosity distributions.  In a
companion paper, \citet{rig09} compare \ofour luminosity to hard
(14--195~keV) X-ray luminosity for the RSA Sy1s, and establish \ofour
to be a measure of intrinsic AGN luminosity.  Thus our result
indicates that the Sy1s and Sy2s in the RSA sample are consistent with
being drawn from the same parent distribution of intrinsic luminosity.
It also confirms that the RSA sample is one of the least-biased AGN
samples known, and is well suited for tests of the unification
paradigm and the nature of the obscuring material around AGNs.
Furthermore, we find that the observed \oiii~$\lambda5007$~\AA\ and
2--10~keV X-ray luminosities are biased indicators of AGN intrinsic
luminosity, confirming the results of \citet{mel08a}.

\subsection{Implications for X-ray-selected AGN samples}

It has been argued from deep X-ray surveys with {\it Chandra}
\citep[e.g.,][]{ued03,ste03} and {\it XMM-Newton}
\citep[e.g.,][]{laf05} that the obscured AGN fraction decreases with
increasing luminosity.  Such a trend with observed 2--10 keV X-ray
luminosity is also seen in Figure~\ref{fig:xray}.  However, we have
shown that the \ofour luminosity distributions of obscured and
unobscured AGNs in the RSA sample are quite similar.  Thus, the trend
in Figure~\ref{fig:xray} is most easily explained as a selection
effect due to obscuration of Sy2s in the 2--10 keV band.  This effect
is quite strong in the RSA sample; while Sy1s constitute only 20\% of
the whole sample, 9/14 of the sources with observed 2--10 keV
luminosities $>10^{42}$~erg~s$^{-1}$ and all three of the sources with
observed $L_X>10^{43}$~erg~s$^{-1}$ are Sy1s.  The RSA does not
include sources at the bright end of the X-ray luminosity function
($\sim10^{45}$~erg~s$^{-1}$) and thus is not able to probe the
luminosity dependence of the obscured AGN fraction to the highest
luminosities, but the large selection effects at lower luminosity are
striking.  As suggested by our results for local AGNs, \citet{dwe06}
find that the absorption of X-ray sources in the Chandra Deep Field
South is best described by models where the obscured fraction is
constant with luminosity, and \citet{tre05} also argue that the the
observed decrease with luminosity can be explained as a selection
effect.  Interestingly, a lower incidence of obscured sources at
higher luminosities is found by \citet{saz07} in an all-sky X-ray with
INTEGRAL in the 17-60~keV band, which covers $\sim75$\% of the sky
down to a relatively shallow flux level
$f=7\times10^{-11}$~erg~cm$^{-2}$~s$^{-1}$.  This result may indicate
that even very hard X-rays are biased tracers of obscured AGNs, as
suggested by \citet{rig09}.  If so, then deep surveys with Chandra and
XMM-Newton will remain biased up to redshifts of $z\sim3$ or more,
despite their sampling of rest-frame X-ray energies $\geq20$~keV at
these redshifts.

\subsection{Possible Missing AGNs}

What sources could the RSA sample be missing?  The most obvious group
of AGNs would be those that lack signs of accretion activity in
optical spectra \citep[e.g.,][]{rig06, gho08}.  It is clear from
Figure~\ref{fig:o3} that there are some objects in the RSA with
extreme $\oiv/\oiii$ ratios, but these objects are all at the bright
end of the \ofour flux distribution with
$f>3\times10^{-13}$~erg~cm$^{-2}$~s$^{-1}$ and corresponding \othree
fluxes $\sim10^{-14}$~erg~cm$^{-2}$~s$^{-1}$.  It is reasonable to
expect that sources with such highly extincted \othree would also
exist at fainter flux levels, but may be missed by optical
emission-line selection.  Such selection would also miss any AGN so
deeply embedded that the continuum source is not able to photoionize
the narrow-line region.  This implies that the 4:1 ratio of obscured
to unobscured Seyferts in the RSA sample is a lower limit.

We have made a preliminary evaluation of the incidence of ``missing''
obscured Seyferts from the {\it Spitzer} spectra of star-forming
galaxies in the SINGS sample \citep{ken03} published by \citet{dal09}.
Among the $B_T\leq13$ RSA galaxies in \citet{dal09}, 28/51 have
$>2\sigma$ \ofour detections, but many of these lines are quite weak,
especially when compared to \neii~$\lambda12.81~\mu$m.  Except for
NGC1705, every galaxy with $\oiv/\neii>0.05$ \citep[corresponding to
$>2\%$ AGN contribution;][]{stu02,arm07} is optically classified as an
AGN (LINER or Seyfert) by J.~Moustakas et~al. (2009, in preparation).
The remaining candidate, NGC1705, is a dwarf starburst galaxy with an
\ofour luminosity $<10^{38}$~erg~s$^{-1}$, smaller than any of the RSA
Seyferts.  Ultraviolet spectra of this galaxy show evidence for
Wolf-Rayet stars \citep{meu92} that could explain its observed
$\oiv/\neii$ ratio \citep[e.g.,][]{sch99}.  It is also anomalously
weak in the radio, even for a star-forming galaxy \citep{can06}, so we
conclude that it is not likely that NGC1705 contains a genuine active
nucleus.  Thus the incidence of AGNs that are missed by optical
emission-line selection does not appear to be large \citep[but see
also][]{sat08}.

\section{Summary and Conclusions}

We have presented measurements of \oiv~$\lambda25.89~\mu$m luminosity
for 89 Seyfert galaxies from the RSA sample and compared the
distributions of \oiv, \oiii~$\lambda5007$~\AA, 2--10~keV X-ray, and
6~cm radio luminosities among Sy1s and Sy2s.  We find that the
distribution of \ofour luminosities for Sy2s is indistinguishable from
that for Sy1s, while their \othree luminosity distributions are
statistically different.  Under the assumption that \ofour is an
accurate tracer of intrinsic AGN luminosity, this indicates that the
obscured and unobscured RSA Seyferts are consistent with being drawn
from the same parent luminosity distribution, and argues against
models where the ratio of obscured to unobscured AGNs depends on
luminosity.  It also indicates that there is significant extinction
towards or within the narrow-line region in a subset of Sy2s.
Additionally, we find that obscured and unobscured AGNs have similar
distributions of radio luminosities, while their observed X-ray
luminosities are quite different, which provides insight into the
nature of the sources missed by X-ray surveys.

\acknowledgments

We acknowledge helpful discussions with Jennifer Donley, Yong Shi,
Suresh Sivanadam, Amelia Stutz, Jonathan Trump, and Benjamin Weiner.
We thank John Moustakas for providing optical classifications for the
SINGS sample in advance of publication and for comments on the
manuscript.  We also thank Miwa Block for assistance with the
reduction of IRS spectral mapping data, Gonzalo Aniano for independent
measurements of the \ofour line fluxes, and the anonymous referee for
valuable feedback.  This work was partially supported by contract
1255094 from JPL/CalTech to the University of Arizona.

{\it Facilities:} \facility{Spitzer}

\clearpage

\LongTables
\begin{deluxetable*}{lrlrrrrrrrrl}
\tabletypesize{\scriptsize}
\tablecaption{Table of Measurements\label{table:data}}
\tablewidth{0pt}
\tablehead{
 &  & \colhead{Seyfert} & & & & \\
\colhead{NAME} & \colhead{D} & \colhead{Type} & \colhead{\oiv} & \colhead{$\sigma$} & \colhead{\oiii} & \colhead{Ref} & \colhead{6~cm} & \colhead{Ref} & \colhead{$2-10$~keV}  & \colhead{$N_H$}  &  \colhead{Ref} \\
 (1) & (2) & (3) & (4) & (5) & (6) & (7) & (8) & (9) & (10) & (11) & (12)}
\startdata
            NGC777  &   66.5  &    2 & $<$4.35e-14  &  1.45e-14  &  $<$2.4e-15  &   1  &     1.4e-26  &  11  &     2.4e-13  &     \nodata  &  29  \\
            NGC788  &   54.1  &    2 &    1.80e-13  &  0.08e-13  &     3.6e-14  &   2  &     1.2e-26  &  12  &     4.6e-12  &     3.0e+23  &  30  \\
           NGC1068  &   14.4  &  1.8 &    1.90e-11\tablenotemark{d}  &  0.19e-11  &     3.6e-12  &   2  &     5.2e-24  &  11  &     3.8e-12 &  $>$1.0e+25  &  31,32  \\
           NGC1058  &    9.2\tablenotemark{a}  &    2 & $<$2.66e-14  &  0.69e-14  &     1.4e-15  &   1  &  $<$1.2e-27  &  11  &  $<$4.0e-14 &     \nodata  &  33  \\
           NGC1097  &   16.5  &  1.0 & $<$3.74e-13  &  0.57e-13  &     1.5e-14  &   3  &     3.8e-26  &  13  &     1.7e-12  &     2.3e+20  &  34  \\
           NGC1241  &   53.8  &    2 & $<$1.40e-13  &  0.16e-13  &     1.6e-15  &   4  &     1.0e-25  &  14  &     \nodata   &     \nodata  &      \\
           NGC1275  &   70.1  &  1.5 & $<$1.85e-13  &  0.40e-13  &     6.9e-13  &   1  &     2.1e-22  &  11  &     1.4e-11  &  $<$1.5e+21  &  35,36  \\
           NGC1365  &   21.5  &  1.8 &    1.58e-12  &  0.12e-12  &     6.2e-14  &   3  &     9.0e-27  &  15  &     5.9e-12  &     4.0e+23  &  37  \\
           NGC1358  &   53.6  &    2 &    7.61e-14  &  1.43e-14  &     2.0e-13  &   1  &     1.2e-26  &  11  &     3.4e-13  &     \nodata  &  30  \\
           NGC1386  &   10.6  &    2 &    8.70e-13  &  0.27e-13  &     2.7e-13  &   3  &     1.3e-25  &  16  &     2.1e-13  &  $>$1.0e+24  &  38  \\
           NGC1433  &   13.3  &    2 &    6.07e-14  &  1.12e-14  &     2.1e-14  &   3  &  $<$3.0e-26  &  16  &     \nodata   &     \nodata  &      \\
           NGC1566  &   19.4  &  1.5 &    8.88e-14  &  0.46e-14  &     1.7e-13  &   3  &  $<$6.0e-26  &  16  &     6.7e-12  &     \nodata  &  39  \\
           NGC1667  &   61.2  &    2 &    9.28e-14  &  1.48e-14  &     9.1e-15  &   2  &     1.2e-26  &  11  &     2.6e-14  &     \nodata  &  40  \\
           NGC2273  &   28.4  &    2 &    1.47e-13  &  0.20e-13  &     2.8e-13  &   1  &     1.4e-25  &  11  &     6.9e-13  &  $>$1.8e+24  &  41  \\
           NGC2639  &   42.6  &  1.9 &    3.27e-14  &  0.38e-14  &     1.9e-14  &   1  &     1.7e-24  &  11  &     8.5e-14  &     \nodata  &  30  \\
           NGC2685  &   16.2  &    2 &    1.15e-14  &  0.35e-14  &     8.1e-15  &   1  &  $<$1.3e-27  &  11  &     2.7e-13  &     \nodata  &  33  \\
           NGC2655  &   24.4  &    2 &    6.25e-14  &  1.41e-14  &     3.9e-14  &   1  &     3.4e-25  &  11  &     1.0e-12  &     4.5e+23  &  42  \\
           NGC2992  &   34.1  &  1.9 &    1.08e-12  &  0.03e-12  &     5.2e-14  &   2  &     7.0e-26  &  16  &     6.3e-12  &     1.4e+22  &  43  \\
           NGC3031  &    3.6\tablenotemark{a}  &  1.5 &    4.99e-14  &  0.94e-14  &     2.2e-13  &   1  &     8.4e-25  &  11  &     1.2e-11 &     1.0e+21  &  33  \\
           NGC3081  &   34.2  &    2 &    9.89e-13  &  0.20e-13  &     2.1e-13  &   2  &     9.0e-27  &  12  &     1.3e-12  &     6.4e+23  &  44  \\
           NGC3079  &   20.4  &    2 &    1.53e-13  &  0.38e-13  &     1.8e-15  &   1  &     9.2e-25  &  11  &     3.7e-13  &  $>$1.0e+25  &  45  \\
            IC2560  &   40.7  &    2 &    5.43e-13  &  0.17e-13  &     1.3e-13  &   2  &     9.5e-26  &  13  &     3.6e-13  &  $>$1.0e+24  &  46,41  \\
           NGC3147  &   40.9  &    2 & $<$6.50e-14  &  1.35e-14  &     1.7e-14  &   1  &     1.0e-25  &  11  &     1.5e-12  &  $<$1.7e+22  &  47,48  \\
           NGC3185  &   21.3  &    2 &    4.70e-14  &  1.62e-14  &     5.0e-14  &   1  &     1.9e-27  &  11  &     2.0e-14  &  $>$1.0e+24  &  33  \\
           NGC3227  &   20.6  &  1.5 &    5.71e-13  &  0.45e-13  &     9.4e-13  &   1  &     2.0e-25  &  11  &     2.3e-11  &     1.9e+22  &  49,50  \\
           NGC3254  &   23.6  &    2 & $<$1.47e-14  &  0.45e-14  &     6.0e-15  &   1  &  $<$1.2e-27  &  11  &  $<$1.2e-14  &     \nodata  &  29  \\
           NGC3281  &   44.7  &    2 &    1.39e-12  &  0.04e-12  &     1.3e-14  &   2  &     2.7e-25  &  12  &     2.9e-12  &     1.5e+24  &  51  \\
           NGC3486  &    7.4  &    2 &    3.30e-14  &  1.16e-14  &     1.3e-14  &   1  &  $<$1.2e-27  &  11  &     5.0e-14  &     \nodata  &  52  \\
           NGC3516  &   38.9  &  1.2 &    5.60e-13  &  0.23e-13  &     3.5e-13  &   1  &     3.2e-26  &  11  &     1.4e-11  &     7.9e+21  &  53  \\
    IRAS11215-2806  &   62.4  &    2 &    9.97e-14  &  0.64e-14  &     5.1e-14  &   5  &     1.8e-25  &  17  &     \nodata   &     \nodata  &      \\
           NGC3735  &   41.0  &    2 &    4.84e-13  &  0.18e-13  &     3.7e-14  &   1  &     8.1e-27  &  11  &  $<$9.2e-14  &     \nodata  &  29  \\
           NGC3783  &   36.1  &  1.2 &    2.80e-13  &  0.25e-13  &     8.3e-13  &   6  &     1.3e-25  &  16  &     7.0e-11  &     8.7e+21  &  30,54  \\
           NGC3941  &   18.9  &    2 &    9.35e-15  &  4.59e-15  &     7.7e-15  &   1  &     2.3e-27  &  11  &     4.0e-14  &     \nodata  &  33  \\
           NGC3976  &   37.7  &    2 & $<$1.01e-13  &  0.20e-13  &     7.7e-15  &   1  &     4.5e-27  &  11  &     8.5e-14  &     \nodata  &  29  \\
           NGC3982  &   17.0  &  1.9 & $<$1.18e-13  &  0.16e-13  &     2.0e-13  &   1  &     1.8e-26  &  11  &  $<$5.0e-14  &  $>$1.6e+24  &  55  \\
           NGC4051  &   17.0  &  1.2 &    2.64e-13  &  0.25e-13  &     4.4e-13  &   1  &     2.1e-26  &  11  &     2.3e-11  &  $<$2.8e+21  &  30,47  \\
           NGC4138  &   17.0  &  1.9 &    4.27e-14  &  0.32e-14  &     1.6e-14  &   1  &     7.8e-27  &  11  &     5.5e-12  &     8.0e+22  &  33  \\
           NGC4151  &   20.3  &  1.5 &    2.08e-12  &  0.08e-12  &     1.1e-11  &   1  &     8.1e-25  &  11  &     4.8e-11  &     3.1e+22  &  56  \\
           NGC4168  &   16.8  &  1.9 &    1.39e-14  &  0.44e-14  &     2.4e-15  &   1  &     5.0e-26  &  11  &  $<$3.6e-14  &     \nodata  &  57  \\
           NGC4235  &   35.1  &  1.2 &    4.33e-14  &  0.78e-14  &     2.0e-14  &   1  &     5.1e-26  &  11  &     1.0e-11  &     3.0e+21  &  30  \\
           NGC4258  &    8.0\tablenotemark{a}  &  1.9 &    7.49e-14  &  1.23e-14  &     1.0e-13  &   1  &     1.8e-26  &  11  &     7.1e-12 &     8.2e+22  &  58  \\
           NGC4378  &   35.1  &    2 & $<$1.83e-14  &  0.61e-14  &     5.0e-15  &   1  &     3.0e-27  &  11  &     1.5e-13  &     \nodata  &  29  \\
           NGC4388  &   16.8  &  1.9 &    2.59e-12  &  0.03e-12  &     1.6e-13  &   2  &     5.4e-26  &  11  &     3.7e-12  &     3.5e+23  &  59  \\
           NGC4395  &    4.6\tablenotemark{b}  &  1.8 &    4.23e-14  &  0.31e-14  &     1.4e-13  &   1  &     6.8e-27  &  11  &     3.7e-12 &     1.2e+22  &  60  \\
           NGC4472  &   16.8  &    2 & $<$6.64e-14  &  1.89e-14  &     1.9e-15  &   1  &     1.9e-25  &  11  &  $<$3.8e-13  &     \nodata  &  33  \\
           NGC4477  &   16.8  &    2 &    1.69e-14  &  0.56e-14  &     1.9e-14  &   1  &     1.7e-27  &  11  &     1.2e-13  &     \nodata  &  33  \\
           NGC4501  &   16.8  &    2 &    3.98e-14  &  0.34e-14  &     3.7e-14  &   1  &     1.1e-26  &  11  &     5.0e-14  &  $>$1.0e+24  &  61  \\
           NGC4507  &   59.6  &    2 &    3.31e-13  &  0.22e-13  &     4.5e-13  &   2  &     1.1e-25  &  18  &     1.8e-11  &     5.9e+23  &  62  \\
           NGC4565  &    9.7  &  1.9 &    2.09e-14  &  1.52e-14  &     1.5e-14  &   1  &     2.6e-26  &  11  &     2.1e-13  &     2.5e+21  &  63  \\
           NGC4579  &   16.8  &  1.9 &    2.83e-14  &  0.62e-14  &     7.8e-14  &   1  &     3.8e-25  &  11  &     5.5e-12  &     3.3e+21  &  48  \\
           NGC4593  &   41.3  &  1.0 &    1.32e-13  &  0.27e-13  &     1.3e-13  &   5  &     1.6e-26  &  19  &     3.0e-11  &     1.6e+21  &  64  \\
           NGC4594  &   20.0  &  1.9 &    2.62e-14  &  0.43e-14  &     4.7e-14  &   1  &     1.2e-24  &  16  &     1.6e-12  &     1.7e+21  &  65  \\
            IC3639  &   35.3  &    2 & $<$3.55e-13  &  0.73e-13  &     4.1e-13  &   4  &     2.0e-25  &  16  &     8.0e-14  &  $>$1.6e+24  &  54  \\
           NGC4639  &   16.8  &  1.0 &    1.54e-14  &  0.43e-14  &     7.5e-15  &   1  &     2.2e-27  &  11  &     5.0e-13  &     7.3e+20  &  66  \\
           NGC4698  &   16.8  &    2 &    2.03e-14  &  0.37e-14  &     1.9e-14  &   1  &     2.6e-27  &  11  &     4.0e-14  &     \nodata  &  33  \\
           NGC4725  &   12.4  &    2 &    1.24e-14  &  0.31e-14  &     2.0e-14  &   1  &  $<$1.7e-27  &  11  &     4.0e-14  &     \nodata  &  33  \\
           NGC4941  &   16.8  &    2 &    1.50e-13  &  0.18e-13  &     1.4e-13  &   2  &     4.3e-26  &  12  &     8.5e-13  &     6.9e+23  &  67  \\
           NGC4939  &   46.6  &    2 &    4.30e-13  &  0.08e-13  &     1.6e-13  &   2  &     7.0e-27  &  20  &     1.4e-12  &  $>$1.0e+25  &  44  \\
           NGC4945  &    4.3  &    2 &    3.00e-13\tablenotemark{e}  &  0.60e-13  &     \nodata   &      &     2.9e-25  &  21  &     4.0e-12  &     5.0e+24  &  30,68  \\
           NGC5005  &   21.3  &    2 &    1.99e-14  &  1.44e-14  &     4.7e-14  &   1  &     2.7e-26  &  20  &     5.1e-13  &     3.0e+22  &  41  \\
           NGC5033  &   18.7  &  1.5 &    1.59e-13  &  0.07e-13  &     5.3e-14  &   1  &     3.0e-26  &  11  &     2.9e-12  &  $<$8.7e+20  &  33,69  \\
           NGC5128  &    4.3  &    2 &    9.89e-13  &  0.79e-13  &     1.2e-14  &   3  &     7.0e-23  &  22  &     3.8e-10  &     1.0e+23  &  70  \\
           NGC5135  &   57.7  &    2 &    5.83e-13  &  0.34e-13  &     3.6e-14  &   2  &     5.9e-25  &  12  &     2.2e-13  &  $>$1.0e+24  &  71  \\
           NGC5194  &    8.4\tablenotemark{c}  &    2 &    2.46e-13  &  0.10e-13  &     1.1e-13  &   1  &     9.8e-27  &  11  &     4.8e-13 &     5.6e+24  &  33,72  \\
           NGC5273  &   21.3  &  1.5 &    3.72e-14  &  1.42e-14  &     1.2e-13  &   1  &     1.0e-26  &  11  &     6.7e-12  &     9.0e+21  &  33  \\
           NGC5395  &   46.7  &    2 & $<$9.29e-14  &  1.77e-14  &     1.8e-15  &   1  &     1.9e-27  &  11  &     \nodata   &     \nodata  &      \\
           NGC5427  &   40.4  &    2 &    2.68e-14  &  0.58e-14  &     5.5e-14  &   2  &     2.5e-26  &  13  &  $<$1.1e-13  &     \nodata  &  54  \\
          CIRCINUS  &    2.9  &    2 &    6.79e-12\tablenotemark{d}  &  1.36e-12  &     8.3e-14  &   7  &     3.4e-25  &  23  &     1.4e-11 &     4.0e+24  &  73  \\
           NGC5506  &   30.0  &  1.9 &    2.22e-12  &  0.07e-12  &     1.6e-13  &   2  &     1.6e-24  &  16  &     6.9e-11  &     3.2e+22  &  74  \\
           NGC5631  &   32.7  &    2 &    1.46e-14  &  0.39e-14  &     4.5e-15  &   1  &     3.7e-27  &  11  &     \nodata   &     \nodata  &      \\
           NGC5643  &   14.4  &    2 &    8.16e-13  &  0.41e-13  &     2.4e-13  &   2  &     1.6e-26  &  24  &     6.3e-13  &  $>$1.0e+24  &  75  \\
           NGC5728  &   41.1  &    2 &    1.29e-12  &  0.02e-12  &     1.2e-13  &   2  &     5.2e-26  &  25  &     1.8e-12  &     8.2e+23  &  76  \\
           NGC5899  &   42.8  &    2 &    2.63e-13  &  0.15e-13  &     6.9e-14  &   8  &     4.0e-26  &  26  &     \nodata   &     \nodata  &      \\
           NGC6221  &   19.3  &    2 & $<$4.62e-13  &  0.90e-13  &     2.5e-14  &   2  &  $<$1.0e-26  &  16  &     1.4e-11  &     1.0e+22  &  77  \\
           NGC6300  &   14.0  &    2 &    2.98e-13  &  0.47e-13  &     2.3e-14  &   2  &     4.2e-26  &  13  &     1.3e-11  &     2.1e+23  &  78  \\
           NGC6814  &   25.6  &  1.5 &    2.13e-13  &  0.21e-13  &     7.0e-14  &   6  &     2.0e-26  &  16  &     1.1e-12  &  $<$5.8e+20  &  79  \\
           NGC6951  &   24.1  &    2 &    8.37e-14  &  2.01e-14  &     7.0e-15  &   1  &     1.0e-26  &  11  &  $<$5.7e-14  &     \nodata  &  29  \\
            MRK509  &  143.8  &  1.2 &    2.85e-13  &  0.11e-13  &     8.6e-13  &   9  &     1.8e-26  &  27  &     4.5e-11  &     2.1e+21  &  80  \\
           NGC7130  &   68.7  &    2 &    1.67e-13  &  0.34e-13  &     1.1e-13  &   2  &     3.8e-25  &  16  &     1.6e-13  &  $>$1.0e+24  &  81  \\
           NGC7172  &   37.6  &    2 &    4.86e-13  &  0.15e-13  &     1.3e-14  &   2  &     1.2e-25  &  13  &     1.1e-11  &     1.1e+23  &  48  \\
           NGC7213  &   24.9  &  1.5 &    2.11e-14  &  1.21e-14  &     3.4e-13  &   6  &     2.1e-24  &  18  &     3.3e-11  &  $<$4.2e+21  &  48  \\
           NGC7314  &   20.8  &  1.9 &    4.91e-13  &  0.12e-13  &     2.2e-14  &   2  &     2.7e-26  &  13  &     2.4e-11  &     9.3e+21  &  48  \\
           NGC7410  &   24.8  &    2 &    4.63e-14  &  1.12e-14  &     1.9e-14  &   4  &     1.4e-26  &  28  &     \nodata   &     \nodata  &      \\
           NGC7469  &   67.0  &  1.2 &    3.67e-13  &  0.86e-13  &     5.9e-13  &  10  &     2.1e-25  &  16  &     3.1e-11  &     1.3e+20  &  30  \\
           NGC7479  &   32.4  &  1.9 & $<$2.67e-13  &  0.62e-13  &     1.1e-14  &   1  &     2.7e-26  &  11  &     1.5e-13  &  $>$1.0e+24  &  29  \\
           NGC7496  &   23.1  &    2 & $<$1.87e-13  &  0.48e-13  &     9.6e-15  &   2  &     5.8e-26  &  14  &     \nodata   &     \nodata  &      \\
           NGC7582  &   22.0  &    2 &    2.22e-12  &  0.16e-12  &     5.7e-14  &   2  &     6.9e-25  &  16  &     7.6e-12  &     2.3e+23  &  82  \\
           NGC7590  &   22.0  &    2 &    6.88e-14  &  1.22e-14  &     1.1e-14  &   2  &  $<$3.0e-27  &  14  &     7.7e-14  &     \nodata  &  29  \\
           NGC7743  &   24.4  &    2 &    3.30e-14  &  1.79e-14  &     7.9e-15  &   2  &     2.8e-26  &  11  &  $<$3.5e-14  &     \nodata  &  29  \\
\enddata
\tablecomments{Col. (1): Galaxy name.  Col. (2): Distance  [Mpc].  Col. (3): Optical classification from \citet{mai95} or \citet{ho97}.  Col. (4): Observed \oiv~$\lambda25.89~\mu$m flux [erg~cm$^{-2}$~s$^{-1}$].  Col. (5): \ofour uncertainty [erg~cm$^{-2}$~s$^{-1}$].  Col. (6): Observed \oiii~$\lambda5007$ flux [erg~cm$^{-2}$~s$^{-1}$].  No extinction corrections have been applied.  Col. (7): Reference for \othree flux.  Col. (8): Observed 6~cm radio flux density [erg~cm$^{-2}$~s$^{-1}$~Hz$^{-1}$].  Col. (9): Reference for 6~cm flux density.  Col. (10): Observed 2--10 keV X-ray flux [erg~cm$^{-2}$~s$^{-1}$].  No extinction corrections have been applied.  Col. (11): Hydrogen column density measured from X-ray observations [cm$^{-2}$].  Col. (12): Reference(s) for X-ray flux and column density.}
\tablecomments{References:  (1) \citet{ho97}.  (2) \citet{gu06}.  (3) \citet{ver86}.  (4) \citet{vac97}.  (5) \citet{deg92}.  (6) \citet{win92}.  (7) \citet{oli94}.  (8) \citet{sta82}.  (9) \citet{cru94}.  (10) \citet{fri89}.  (11) \citet{ho01}.  (12) \citet{ulv89}.  (13) \citet{mor99}.  (14) \citet{the00}.  (15) \citet{san95}.  (16) \citet{sad95}.  (17) \citet{sch01}.  (18) \citet{bra98}.  (19) \citet{ulv84}.  (20) \citet{vil90}.  (21) \citet{jon94}.  (22) \citet{sle94}.  (23) \citet{dav98}.  (24) \citet{kew00}.  (25) \citet{sch88}.  (26) \citet{whi97}.  (27) \citet{nef92}.  (28) \citet{con98}.  (29) This work.  (30) \citet{tur01}.  (31) \citet{you01}.  (32) \citet{mat00}.  (33) \citet{cap06}. (34) \citet{nem06}.  (35) \citet{eva06}.  (36) \citet{bal06b}.  (37) \citet{ris05}.  (38) \citet{lev06}.  (39) \citet{sax08}.  (40) \citet{tur97}.  (41) \citet{gua05b}.  (42) \citet{ter02}.  (43) \citet{gil00}.  (44) \citet{mai98}.  (45) \citet{iyo01}.  (46) \citet{iwa02}.  (47) \citet{bia08}.  (48) \citet{dad07}.  (49) \citet{yaq04}.  (50) \citet{mck07}.  (51) \citet{vig02}.  (52) \citet{pap01}.  (53) \citet{net02}.  (54) \citet{kas01}.  (55) \citet{gua05}.  (56) \citet{yan01}.  (57) \citet{bal06}.  (58) \citet{yan07}.  (59) \citet{iwa03}.  (60) \citet{mor05}.  (61) \citet{bri08}.  (62) \citet{ris02}.  (63) \citet{chi06}.  (64) \citet{stee03}.  (65) \citet{pel02}.  (66) \citet{ho99}.  (67) \citet{sal97}.  (68) \citet{don96}.  (69) \citet{ter99}.  (70) \citet{gra03}.  (71) \citet{lev04}.  (72) \citet{fuk01}.  (73) \citet{bia02}.  (74) \citet{bia04}.  (75) \citet{bia06}.  (76) \citet{zha06}.  (77) \citet{lev01}.  (78) \citet{gua02}.  (79) \citet{rey97}.  (80) \citet{yaq03}.  (81) \citet{lev05}.  (82) \citet{don04}.    }
\tablenotetext{a}{Cepheid distance from \citet{fre01}.}
\tablenotetext{b}{Tip of the red giant branch distance from \citet{kar03}.}
\tablenotetext{c}{Planetary nebula luminosity function distance from \citet{fel97}.}
\tablenotetext{d}{Flux from \citet{stu02}.  A 20\% calibration uncertainty is adopted.}
\tablenotetext{e}{Flux from \citet{spo00}.  A 20\% calibration uncertainty is adopted.}
\end{deluxetable*}

\vskip1.5cm

\begin{deluxetable}{lcccc}
\tabletypesize{\scriptsize}
\tablecaption{\ofour Statistical Tests\label{table:oiv_stats}}
\tablewidth{0pt}

\tablehead{ \colhead{Sy1 v. } & \colhead{Sy2} & \colhead{Sy2,
Compton-thin} & \colhead{Sy2, Compton-thick} & \colhead{Sy2,
$N_H$ unknown}\\ \colhead{(1)} & \colhead{(2)} & \colhead{(3)} &
\colhead{(4)} & \colhead{(5)}}

\startdata
Gehan, permutation\tablenotemark{a}     & 0.322 & 0.979 & 0.680 & 0.010 \\
Gehan, hypergeometric\tablenotemark{b}  & 0.309 & 0.979 & 0.682 & 0.007 \\
logrank\tablenotemark{c}                & 0.464 & 0.960 & 0.570 & 0.040 \\
Peto-Peto\tablenotemark{d}              & 0.289 & 0.963 & 0.693 & 0.010 \\
Peto-Prentice\tablenotemark{e}          & 0.298 & 0.966 & 0.696 & 0.009 \\
\enddata

\tablecomments{Col. (1): Seyfert types 1.0--1.5.  18 objects, 2 upper
limits.  Col. (2): Seyfert types 1.8--2.  71 objects, 14 upper limits.
Col. (3): Seyfert types 1.8--2, $N_H<1\times10^{-24}$~cm$^{-2}$.  24
objects, 2 upper limits.  Col. (4): Seyfert types 1.8--2,
$N_H>1\times10^{-24}$~cm$^{-2}$.  18 objects, 3 upper limits.
Col. (5): Seyfert types 1.8--2, $N_H$ unknown.  29 objects, 9 upper
limits.  The values in Columns 2--5 correspond to probabilities that
the two samples are drawn from the same distribution.}

\tablenotetext{a}{Gehan's generalized Wilcoxon test, permutation
variance \citep{fei85,lav92}.}

\tablenotetext{b}{Gehan's generalized Wilcoxon test, hypergeometric
variance \citep{fei85,lav92}.} 

\tablenotetext{c}{\textnormal{logrank test} \citep{fei85,lav92}.}

\tablenotetext{d}{Peto \& Peto generalized Wilcoxon test
\citep{fei85,lav92}.}

\tablenotetext{e}{Peto \& Prentice generalized Wilcoxon test
\citep{fei85,lav92}.}
\end{deluxetable}

\clearpage

\begin{deluxetable}{lcccc}
\tabletypesize{\scriptsize}
\tablecaption{\othree Statistical Tests\label{table:o3_stats}}
\tablewidth{0pt}

\tablehead{ \colhead{Sy1 v. } & \colhead{Sy2} & \colhead{Sy2,
Compton-thin} & \colhead{Sy2, Compton-thick} & \colhead{Sy2, $N_H$
unknown}\\ \colhead{(1)} & \colhead{(2)} & \colhead{(3)} &
\colhead{(4)} & \colhead{(5)}}

\startdata
Gehan, permutation     & 0.002            & 0.015   & 0.166   & $3\times10^{-4}$  \\
Gehan, hypergeometric  & $2\times10^{-4}$ & 0.015   & 0.164   & $6\times10^{-5}$  \\
logrank                & 0.004            & 0.003   & 0.054   & $6\times10^{-4}$  \\
Peto-Peto              & 0.002            & 0.003   & 0.054   & $3\times10^{-4}$  \\
Peto-Prentice          & $6\times10^{-4}$ & \nodata & \nodata & $2\times10^{-4}$  \\
\enddata

\tablecomments{Col. (1): Seyfert types 1.0--1.5.  18 objects, 0 upper
limits.  Col. (2): Seyfert types 1.8--2.  71 objects.  1 upper limit.
1 object with no data (NGC4945).  Col. (3): Seyfert types 1.8--2,
$N_H<1\times10^{-24}$~cm$^{-2}$.  24 objects, 0 upper limits.
Col. (4): Seyfert types 1.8--2, $N_H>1\times10^{-24}$~cm$^{-2}$.  18
objects, 0 upper limits, 1 object with no data (NGC4945).  Col. (5):
Seyfert types 1.8--2, $N_H$ unknown.  29 objects, 1 upper limit.  The
values in Columns 2--5 correspond to probabilities that the two
samples are drawn from the same distribution.  The Peto \& Prentice
Wilcoxon test reduces to Gehan's Wilcoxon test when there are no upper
limits.}

\end{deluxetable}
\vskip1.5cm

\begin{deluxetable}{lcccc}
\tabletypesize{\scriptsize}
\tablecaption{6~cm Statistical Tests\label{table:radio_stats}}
\tablewidth{0pt}

\tablehead{ \colhead{Sy1 v. } & \colhead{Sy2} & \colhead{Sy2,
Compton-thin} & \colhead{Sy2, Compton-thick} & \colhead{Sy2, $N_H$
unknown}\\ \colhead{(1)} & \colhead{(2)} & \colhead{(3)} &
\colhead{(4)} & \colhead{(5)}}

\startdata
Gehan, permutation      & 0.191 & 0.682 & 0.786 & 0.023 \\
Gehan, hypergeometric   & 0.172 & 0.682 & 0.787 & 0.017 \\
logrank                 & 0.078 & 0.517 & 0.677 & 0.006 \\
Peto-Peto               & 0.188 & 0.682 & 0.774 & 0.021 \\
Peto-Prentice           & 0.197 & 0.685 & 0.775 & 0.021 \\

\enddata

\tablecomments{Col. (1): Seyfert types 1.0--1.5.  18 objects, 1 upper
limit.  (2): Seyfert types 1.8--2.  71 objects, 8 upper limits.
Col. (3): Seyfert types 1.8--2, $N_H<1\times10^{-24}$~cm$^{-2}$.  24
objects, 1 upper limit.  Col. (4): Seyfert types 1.8--2,
$N_H>1\times10^{-24}$~cm$^{-2}$.  18 objects, 0 upper limits.
Col. (5): Seyfert types 1.8--2, $N_H$ unknown.  29 objects, 7 upper
limits.  The values in Columns 2--5 correspond to probabilities that
the two samples are drawn from the same distribution.}

\end{deluxetable}

\vskip1.5cm

\begin{deluxetable}{lcccc}
\tabletypesize{\scriptsize}
\tablecaption{2--10~keV Statistical Tests\label{table:xray_stats}}
\tablewidth{0pt}

\tablehead{ \colhead{Sy1 v. } & \colhead{Sy2} & \colhead{Sy2,
Compton-thin} & \colhead{Sy2, Compton-thick} & \colhead{Sy2, $N_H$
unknown}\\ \colhead{(1)} & \colhead{(2)} & \colhead{(3)} &
\colhead{(4)} & \colhead{(5)}}

\startdata
Gehan, permutation      & $1\times10^{-5}$  & 0.057   & $3\times10^{-5}$  & $<1\times10^{-7}$ \\
Gehan, hypergeometric   & $<1\times10^{-7}$ & 0.058   & $9\times10^{-6}$  & $<1\times10^{-7}$ \\
logrank                 & $<1\times10^{-7}$ & 0.015   & $3\times10^{-6}$  & $<1\times10^{-7}$ \\
Peto-Peto               & $1\times10^{-5}$  & 0.015   & $3\times10^{-5}$  & $<1\times10^{-7}$ \\
Peto-Prentice           & $3\times10^{-6}$  & \nodata & $1\times10^{-5}$  & $<1\times10^{-7}$ \\

\enddata

\tablecomments{Col. (1): Seyfert types 1.0--1.5.  18 objects, 0 upper
limits.  Col. (2): Seyfert types 1.8--2.  71 objects, 9 upper limits.
8 objects with no data.  Col. (3): Seyfert types 1.8--2,
$N_H<1\times10^{-24}$~cm$^{-2}$.  24 objects, 0 upper limits.
Col. (4): Seyfert types 1.8--2, $N_H>1\times10^{-24}$~cm$^{-2}$.  18
objects, 1 upper limit.  Col. (5): Seyfert types 1.8--2, $N_H$
unknown.  29 objects, 8 upper limits, 8 objects with no data.  The
values in Columns 2--5 correspond to probabilities that the two
samples are drawn from the same distribution.  The Peto \& Prentice
Wilcoxon test reduces to Gehan's Wilcoxon test when there are no upper
limits.}

\end{deluxetable}

\end{document}